\documentclass[%
reprint,
superscriptaddress,
preprintnumbers,
 amsmath,amssymb,
 aps,
prl,
floatfix,
]{revtex4-2}

\usepackage{graphicx}
\usepackage{dcolumn}
\usepackage{bm}

\usepackage{textcomp}   
\usepackage{multirow}
\usepackage{epsfig,dcolumn}
\usepackage{hyperref}
\usepackage{comment}
\usepackage{subfigure}  
\begin{document}

\title{First glimpse of the $N=82$ shell closure below $Z=50$\\ from masses of neutron-rich cadmium isotopes and isomers}

\author{V. Manea}
\email[Corresponding author: ]{vladimir.manea@cern.ch}
\affiliation{CERN, 1211 Geneva 23, Switzerland}
\affiliation{Max-Planck-Institut f\"{u}r Kernphysik, 69117 Heidelberg, Germany}
\affiliation{Instituut voor Kern- en Stralingsfysica, Katholieke Universiteit Leuven, B-3001 Leuven, Belgium}

\author{J.$\>$Karthein$\>$}
\altaffiliation{This article contains data from the Ph.D thesis work of Jonas Karthein, enrolled at the {Ruprecht-Karls-Universit\"at} Heidelberg}
\affiliation{CERN, 1211 Geneva 23, Switzerland}
\affiliation{Max-Planck-Institut f\"{u}r Kernphysik, 69117 Heidelberg, Germany}
    
\author{D. Atanasov}
\altaffiliation[Present address: ]{CERN, 1211 Geneva 23, Switzerland}
\affiliation{Technische Universit\"{a}t Dresden, 01069 Dresden, Germany}

\author{M. Bender}
\affiliation{IP2I Lyon, CNRS/IN2P3, Universit{\'e} de Lyon, Universit{\'e} Claude Bernard Lyon 1, F-69622, Villeurbanne, France}
\author{K. Blaum}
\affiliation{Max-Planck-Institut f\"{u}r Kernphysik, 69117 Heidelberg, Germany}

\author{T. E. Cocolios}
\affiliation{Instituut voor Kern- en Stralingsfysica, Katholieke Universiteit Leuven, B-3001 Leuven, Belgium}

\author{S.$\>$Eliseev}
    \affiliation{Max-Planck-Institut f\"{u}r Kernphysik, 69117 Heidelberg, Germany}


\author{A. Herlert}
\affiliation{FAIR GmbH, 64291 Darmstadt, Germany}

\author{J.~D.~Holt}
\affiliation{TRIUMF, 4004 Wesbrook Mall, Vancouver, BC V6T 2A3, Canada}

\author{W. J. Huang}
\altaffiliation[Present address: ]{Max-Planck-Institut f\"{u}r Kernphysik, 69117 Heidelberg, Germany}
\affiliation{CSNSM-IN2P3-CNRS, Universit\'{e} Paris-Sud, 91406 Orsay, France}

\author{Yu. A. Litvinov}
\affiliation{GSI Helmholtzzentrum f\"{u}r Schwerionenforschung GmbH, 64291 Darmstadt, Germany}

\author{D. Lunney}
\affiliation{CSNSM-IN2P3-CNRS, Universit\'{e} Paris-Sud, 91406 Orsay, France}

\author{J.~Men\'endez}
\affiliation{Center for Nuclear Study,
The University of Tokyo, 113-0033 Tokyo, Japan}
\affiliation{Department de F\'isica Qu\`antica i Astrof\'isica, Universitat de Barcelona, 08028 Barcelona, Spain}


\author{M. Mougeot}
\altaffiliation[Present address: ]{CERN, 1211 Geneva 23, Switzerland}
\affiliation{CSNSM-IN2P3-CNRS, Universit\'{e} Paris-Sud, 91406 Orsay, France}

\author{D. Neidherr}
\affiliation{GSI Helmholtzzentrum f\"{u}r Schwerionenforschung GmbH, 64291 Darmstadt, Germany}

\author{L. Schweikhard}
\affiliation{Institut f\"{u}r Physik, Universit\"{a}t Greifswald, 17487 Greifswald, Germany}

\author{A.~Schwenk}
\affiliation{Institut f\"ur Kernphysik, Technische Universit\"at Darmstadt, 64289 Darmstadt, Germany}
\affiliation{ExtreMe Matter Institute EMMI, GSI Helmholtzzentrum f\"ur Schwerionenforschung GmbH, 64291 Darmstadt, Germany}
\affiliation{Max-Planck-Institut f\"{u}r Kernphysik, 69117 Heidelberg, Germany}

\author{J.~Simonis}
\affiliation{\mbox{Institut f\"ur Kernphysik and PRISMA Cluster of Excellence, Johannes Gutenberg-Universit\"at, 55099 Mainz, Germany}}
\affiliation{Institut f\"ur Kernphysik, Technische Universit\"at Darmstadt, 64289 Darmstadt, Germany}
\affiliation{ExtreMe Matter Institute EMMI, GSI Helmholtzzentrum f\"ur Schwerionenforschung GmbH, 64291 Darmstadt, Germany}


\author{A. Welker}
\affiliation{CERN, 1211 Geneva 23, Switzerland}
\affiliation{Technische Universit\"{a}t Dresden, 01069 Dresden, Germany}

\author{F. Wienholtz}
\altaffiliation[Present address: ]{Institut f\"ur Kernphysik, Technische Universit\"at Darmstadt, 64289 Darmstadt, Germany}
\affiliation{CERN, 1211 Geneva 23, Switzerland}
\affiliation{Institut f\"{u}r Physik, Universit\"{a}t Greifswald, 17487 Greifswald, Germany}
 
\author{K. Zuber}
\affiliation{Technische Universit\"{a}t Dresden, 01069 Dresden, Germany}


\begin{abstract}

We probe the $N=82$ nuclear shell closure by mass measurements of neutron-rich cadmium isotopes with the ISOLTRAP spectrometer at ISOLDE-CERN. The new mass of $^{132}$Cd offers the first value of the $N=82$, two-neutron shell gap below $Z=50$ and confirms the phenomenon of mutually enhanced magicity at $^{132}$Sn. Using the recently implemented phase-imaging ion-cyclotron-resonance method, the ordering of the low-lying isomers in $^{129}$Cd and their energies are determined. The new experimental findings are used to test large-scale shell-model, mean-field and beyond-mean-field calculations, as well as the \textit{ab initio} valence-space in-medium similarity renormalization group.  

\end{abstract}


\date{\today}

\maketitle

The so-called magic numbers of protons and neutrons are associated with large energy gaps in the effective single-particle spectrum of the nuclear mean field \cite{Mayer1955}, revealing shell closures. As such, they are intimately connected to the nuclear interaction and represent essential benchmarks for nuclear models.

Experiments with light radioactive beams have shown that shell closures at $N = 8,20$ and 28 are substantially weakened when the number of protons in the nuclear system is reduced (see \cite{Sorlin2008,Kanungo2013} for a review). New, but weaker shell closures have also been found, e.g., $N=32$ and 34 \cite{Huck85,Wienholtz13,Steppenbeck13,Michimasa2018}. In the shell model, this evolution results from the interplay between the monopole part of the valence-space nucleon-nucleon interaction that determines the single-particle spectrum and multipole forces that induce correlations \cite{Caurier05}. Starting from realistic nuclear forces, the study of closed-shell nuclei provides benchmarks for microscopic calculations of valence-space Hamiltonians, with their many-body contributions \cite{Otsuka10,Hebeler2015,Hagen16,Stro17ENO,Morris2018}. Despite extensive work, significantly less is known for heavier nuclei, in particular for the magic $N = 82$. 



The doubly magic nature of $^{132}$Sn (with 50 protons and 82 neutrons) was reconfirmed recently \cite{Rosiak2018, Gorges2019}. But below $Z=50$ the orbitals occupied by the Fermi-level protons change, as does the proton-neutron interaction, which drives shell evolution. 
This means that without data for nuclides with $Z<50$ and $N \approx 82$, any predictions for the $N=82$ shell gap are rather uncertain. While decay-spectroscopy  \cite{Taprogge2014,Taprogge2015,Lorenz2019}, laser-spectroscopy \cite{Yordanov2013} and mass-spectrometry \cite{Atanasov2015, Lascar2017} studies have been performed for the neutron-rich cadmium isotopes, the energies of the low-lying isomers in $^{129}$Cd and the $N=82$ two-neutron shell gap remain unknown.


 
The $A \approx 130$ $r$-process abundance peak has long been considered an indication of a persistent $N=82$ shell gap in various models. However,  recent studies of $r$-process nucleosynthesis have underlined the importance of fission recycling in certain scenarios, in which the $A=130$ abundance peak is primarily determined by the fission-fragment distribution of $r$-process actinides \cite{Goriely2013,Martin2016}.  

In this work, we present the first direct determination of the $N = 82$ shell gap for $Z<50$ with mass measurements of exotic cadmium isotopes and isomers between $^{124}$Cd and $^{132}$Cd. We exploit all mass-measurement techniques of the ISOLTRAP spectrometer, including the phase-imaging ion-cyclotron-resonance (PI-ICR) method \cite{Eliseev2013, Eliseev14,Karthein2019}. The data are interpreted in comparison to the large-scale shell model and to new calculations made with a beyond-mean-field (BMF) approach \cite{Bender06,Bender08}, as well as the \textit{ab initio} valence-space in-medium similarity renormalization group (VS-IMSRG) \cite{Tsuk12SM,Bogn14SM,Morr15Magnus,Stro16TNO,Stro17ENO,Stro2019ARNPS}. 

The cadmium isotopes were produced at CERN's ISOLDE facility \cite{Borge18} by neutron-induced fission in a uranium-carbide target. The neutrons were produced by 1.4-GeV protons accelerated by CERN's Proton Synchrotron Booster and impinging on a tungsten rod, which reduced contaminants from proton-induced reactions \cite{Koster02}. The neutral products diffused from the $\approx 2000^{\circ}$C target into a hot tantalum cavity where the resonance-ionization laser ion source \cite{Fedosseev12} was used to produce singly charged cadmium ions. A cold quartz line 
\cite{Bouquerel08} greatly suppressed surface ionized cesium and barium contaminants. 

The beam was accelerated to 50~keV, mass separated by the ISOLDE High Resolution Separator and transported to ISOLTRAP for 
accumulation in a segmented, linear radiofrequency quadrupole cooler and buncher \cite{Herfurth01}.  The ion bunch 
was then injected into the multi-reflection time-of-flight mass spectrometer (MR-ToF MS) \cite{Wolf13a} where the cadmium ions were separated from contaminants with a resolving power of $\approx 10^5$. The separated ions were either detected using a secondary electron multiplier for mass measurements, or purified \cite{Wienholtz17} and transported to a tandem Penning-trap system, composed of a preparation trap for beam cooling and further purification \cite{RaimbaultHartmann97,Savard91} and a precision trap for measurements.


\begin{figure}[ht!]
\begin{center}
\includegraphics[width=\columnwidth]{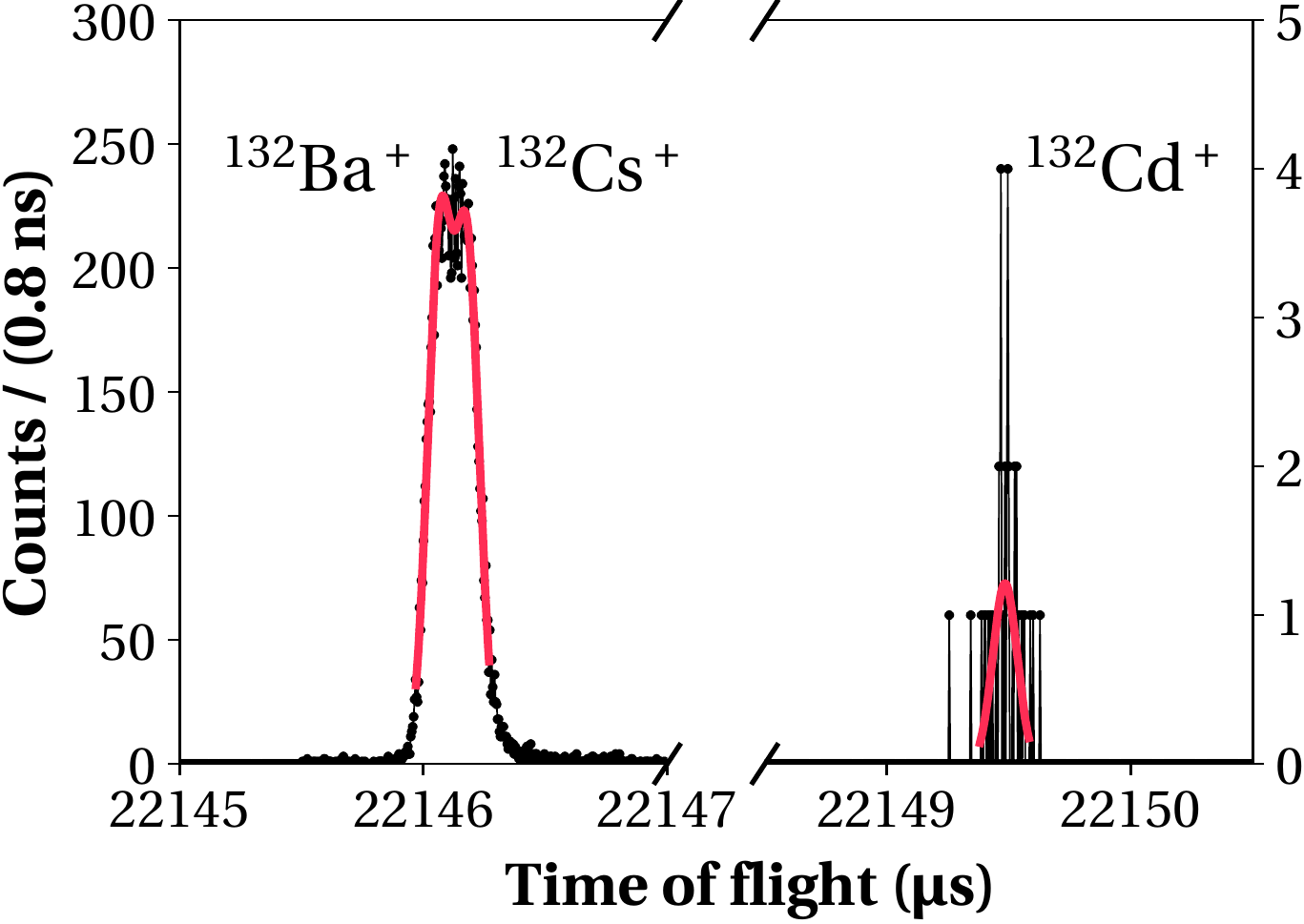}
\caption{MR-ToF spectrum (after 800 revolutions) of $^{132}$Cd$^+$ along with isobaric ions ($^{132}$Ba$^+$ and $^{132}$Cs$^+$), with fits (in red) to Gaussian line shapes. 
}
\label{MRTOF}
\end{center}
\end{figure}
In this work the masses of $^{131,132}$Cd were determined with the MR-ToF MS (see Fig.~\ref{MRTOF}) using a two-parameter calibration formula and hence requiring two reference measurements, as described in \cite{Wienholtz13}. Its short measurement time  of only about 27~ms and direct ion counting made it the method of choice for the most exotic isotopes. Considering only singly charged ions, the mass $m_{i,x}$ of the ion of interest is related to the masses $m_{i,1}$ and $m_{i,2}$ of two reference ions by $m_{i,x}^{1/2} = C_{ToF} \Delta_{Ref} + \frac{1}{2}\Sigma_{Ref}$, with $\Delta_{Ref} = m_{i,1}^{1/2} - m_{i,2}^{1/2}$, $\Sigma_{Ref} = m_{i,1}^{1/2} + m_{i,2}^{1/2}$ and $C_{ToF} = (2t_x - t_1 -t_2)~/~[2(t_1-t_2)]$. The quantities $t_1$, $t_2$ and $t_x$ are the TOFs, measured in the same conditions, of the ions of mass $m_{i,x}$, $m_{i,1}$ and $m_{i,2}$, respectively, with $m_{i,1}$ an isobar of the ion of interest.

The masses of the other studied cadmium isotopes were determined with the precision Penning trap, allowing typically a higher precision and resolving power than the MR-ToF MS, by measuring their cyclotron frequency (as singly charged ions) in the trap, $\nu_{c,x} = q B / (2 \pi m_{i,x})$ (where $q$ is in our case the elementary charge and $B$ is the trap's magnetic-field induction) \cite{Blaum06a}. The atomic mass $m_{x}$ can then be determined as $m_x=r_{ref,x}(m_{ref}-m_e) + m_e$, where $m_e$ is the electron mass and $r_{ref,x} = \nu_{c,ref} / \nu_{c,x}$ is the measured cyclotron-frequency ratio between a singly charged reference ion of atomic mass $m_{ref}$ and the ion of interest. The binding energy of the electron, neglected in the atomic-mass formula, is orders of magnitude smaller than the statistical uncertainty.


Penning-trap measurements of $^{124,126,128,131}$Cd were performed with the time-of-flight ion-cyclotron-resonance (ToF-ICR) method \cite{Konig95a}, including Ramsey-type excitations \cite{George07a,George07b}. 

\begin{figure}[ht!]
\centering
    \includegraphics[width=\linewidth]{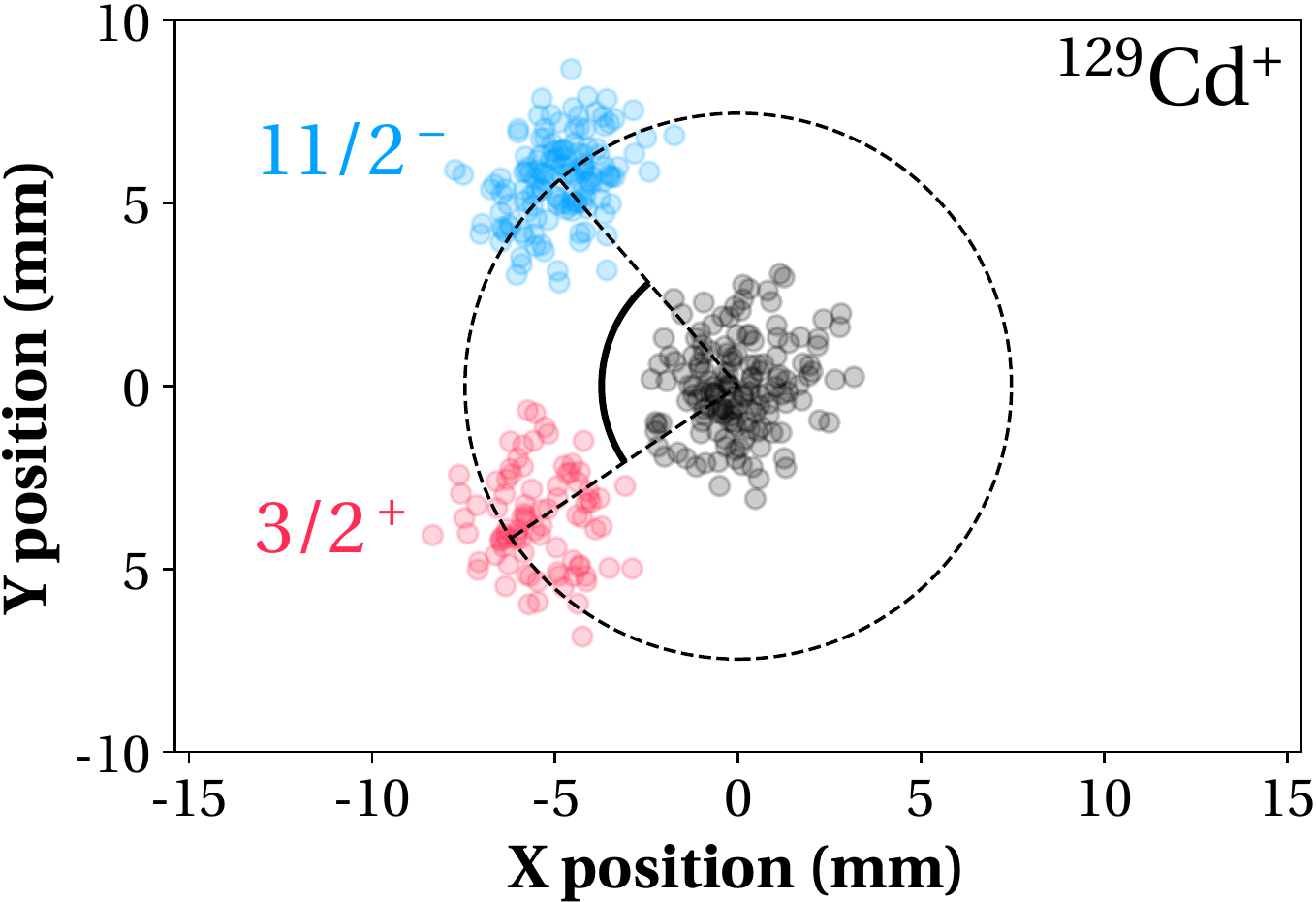}
    \caption{PI-ICR ion projection image of $^{129}$Cd$^+$ with center ion spot measured separately (in black) and the $11/2^-$ (blue) and $3/2^+$ states (red) separated by the marked angle after 106-ms phase accumulation at the modified cyclotron frequency. 
    }
    \label{fig:129Cd}
\end{figure}

For $^{127,129}$Cd the beam was a mixture of ground and isomeric state ($J=3/2^+$ and $J=11/2^-$) which in a prior attempt could not be separated by a long-excitation ToF-ICR measurement \cite{Atanasov2015} due to the short half-lives. In this work we used instead the recently developed PI-ICR method \cite{Eliseev2013,Eliseev14}, by which a radial frequency 
is determined from the phase ``accumulated'' by the circular ion motion in the trap in a given time $t_{\textrm{acc}}$, using its projection on a position-sensitive microchannel-plate detector (MCP). 
In PI-ICR MS one performs three ion-position measurements:  (1) the center of the radial ion trajectory by ejection without preparing a radial motion; (2) for ions prepared on a cyclotron orbit (at frequency $\nu_+$) after evolving for $t_{\textrm{acc}}$; (3) for ions prepared on a magnetron orbit (at frequency $\nu_-$),  after evolving for the same $t_{\textrm{acc}}$. The cyclotron frequency is then given by $\nu_c = [2\pi (n_+ + n_-)+\phi]/(2 \pi t_{\textrm{acc}})$,
where $n_+$ and $n_-$ are the number of integer rotations performed by the ions in steps (2) and (3), respectively, while $\phi$ is the angle between the ion positions measured in the two steps \cite{Eliseev2013, Eliseev14}.

In the second step of the PI-ICR measurement, a resolving power of about $2\times10^6$ was achieved in only $106\,$ms, allowing a clear separation of the two states as illustrated in  Fig.~\ref{fig:129Cd} for $^{129}$Cd$^+$. Their individual masses could thus be determined. 

The experimental results of this work are summarized in Table~\ref{ResTable}.
During the $^{132}$Cd measurements the yield of (stable) $^{132}$Ba$^+$ remained constant, while a gradual increase in the yield of (radioactive) $^{132}$Cs$^+$  was observed.
The data set for $^{132}$Cd was thus split, depending on which isobaric reference dominated, resulting in two independent $C_{TOF}$ values. In this case, as well as for $^{131}$Cd, the weighted averages of the new mass-excess values are used for the figures.

\begin{table*}[ht!]
\centering
\caption{Frequency ratio ($r = \nu_{c,ref} / \nu_{c}$), time-of-flight ratio ($C_{ToF}$) and
mass excess of the cadmium isotopes measured in this work. Mass excesses from the literature
(\cite{Lascar2017} for $^{127}$Cd, \cite{Atanasov2015} for $^{129}$Cd and AME2016 \cite{AME2016} for the rest) are given as well (\# indicates extrapolated values). 
The masses of the reference ions used in the evaluation are from AME2016 \cite{AME2016}.
Experimental half-lives are taken from \cite{NUBASE2016} (and \cite{Lorenz2019} for $^{127}$Cd). The yields, where available, are order-of-magnitude estimates of ion intensities on the ISOLDE central beam line. Values between parentheses are total (statistical plus systematic) uncertainties. 
}
\begin{tabular*}{\textwidth}{c@{\extracolsep{\fill}} c c c c c c c c}
\hline\hline
    \multirow{2}{*}{A} & \multirow{2}{*}{J$^\pi$} & Half-life & Yield &\multirow{2}{*}{Method} & \multirow{2}{*}{References} & \multirow{2}{*}{Ratio r or $C_{ToF}$} & \multicolumn{2}{c}{Mass excess (keV)}\\ 
      &  & (s)  & (Ions/s)  &   & &  & this work   & literature   \\ 
\hline
      	124			 &  $0^+$     &1.25(2)&	 & ToF-ICR &$^{133}$Cs$^+$  	& $r$ = 0.9323743186(432)		& $-76692.4(5.4)$  & $-76701.7(3.0)$ \\ 	
\hline
      	126			 & $0^+$      &0.513(6)&	 &ToF-ICR &$^{133}$Cs$^+$  	& $r$ = 0.9474585581(503)		& $-72249.8(6.2)$  & $-72256.8(2.5)$ \\ 	
\hline
      	\multirow{2}{*}{127}			 &  $3/2^+$      &0.45($^{12}_8$) & $5\times10^4$	 & \multirow{2}{*}{PI-ICR} &\multirow{2}{*}{$^{133}$Cs$^+$}   	& $r$ = 0.9550111122(922)		& $-68737(11)$  & $-68743.4(5.6)$ \\ 

      				 &  $11/2^-$      & 0.36(4) & $1\times10^5$ & 	& & $r$ = 0.9550133972(435)		& $-68453.8(5.4)$  & $-68460.1(4.7)$ \\
\hline
      	128			 &  $0^+$     &0.246(2)&$8\times 10^4 $	 & ToF-ICR & $^{133}$Cs$^+$  	& $r$ = 0.962547502(114)		& $-67225(14)$  & $-67242(7)$ \\ 
\hline
      	\multirow{2}{*}{129}			 & $11/2^-$      & 0.152(6) & $1\times10^4$ & \multirow{2}{*}{PI-ICR} &\multirow{2}{*}{$^{133}$Cs$^+$}  	& $r$ = 0.9701048175(432)		& $-63122.1(5.4)$  & \multirow{2}{*}{$-63058(17)$} \\ 	
      				 & $3/2^+$      & 0.147(3)	 & $5\times10^3$ &  &	& $r$ = 0.9701075886(450)		& $-62779.1(5.6)$  &  \\ 
\hline
	\multirow{2}{*}{131}				 & \multirow{2}{*}{$7/2^-$}      &\multirow{2}{*}{0.098(2)}	 &  \multirow{2}{*}{$3\times10^2$} & ToF-ICR& $^{133}$Cs$^+$  	& $r$ = 0.985217426(252)		& $-55167(31)$  & $-55220(100)$ \\
					 &       &	 & & MR-ToF MS& $^{131}$Cs$^+$,$^{133}$Cs$^+$  	& $C_{ToF} = $ 0.4823166(126)		&  $-55238(24)$ & \\ 
\hline
	\multirow{2}{*}{132}		 & \multirow{2}{*}{$0^+$}      & \multirow{2}{*}{0.082(4)}	 & \multirow{2}{*}{5}& \multirow{2}{*}{MR-ToF MS} & $^{132}$Ba$^+$,$^{133}$Cs$^+$  	& $C_{ToF} = $ 0.4592156(773) 		& $-50499(72)$  & $-50260$\#\\
					 &       &	 & & & $^{132}$Cs$^+$,$^{133}$Cs$^+$  	& $C_{ToF} = $ 0.460420(118) 		& $-50386(110)$  & \\
\hline\hline
 \end{tabular*}
\label{ResTable}
\end{table*}

The analysis of the ToF-ICR measurements followed the procedure in \cite{Kellerbauer03}. For the MR-ToF MS spectra, Gaussian distributions were fit to the data (double-Gaussian for the $^{132}$Ba$^+$/$^{132}$Cs$^+$ double peak) by the binned maximum-likelihood method. When statistically significant, shifts of the $C_{ToF}$ values from changing the fit range, data binning and number of ions simultaneously stored in the MR-ToF MS were included in the total uncertainty.


For the PI-ICR measurements, the unbinned maximum-likelihood fit of the ion-spot positions was performed using 2D Gaussian distributions. The effect of the number of ions simultaneously stored in the trap was studied and, for the analysed data set, was within statistical uncertainties. The mass-dependent shift and systematic uncertainty from \cite{Kellerbauer03} were quadratically added to the total uncertainty. 


The spin assignments for the measured states in $^{127}$Cd and $^{129}$Cd are based on the fact that the high-spin isomers were systematically produced with higher yields, 
corroborated by a laser-spectroscopy study of cadmium isotopes performed at ISOLDE \cite{Yordanov2013} with the same production mechanism, where the yield ratios were determined for $^{127,129}$Cd. We conclude that the excited $11/2^-$ state in $^{127}$Cd becomes the ground state in $^{129}$Cd. The 283(12)-keV excitation energy obtained for $^{127}$Cd agrees with the TITAN result using highly charged ions \cite{Lascar2017}. The 343(8)-keV excitation energy of the $3/2^+$ state in $^{129}$Cd is a new value. 

\begin{figure}[ht!]
\centering
    \includegraphics[width=\linewidth]{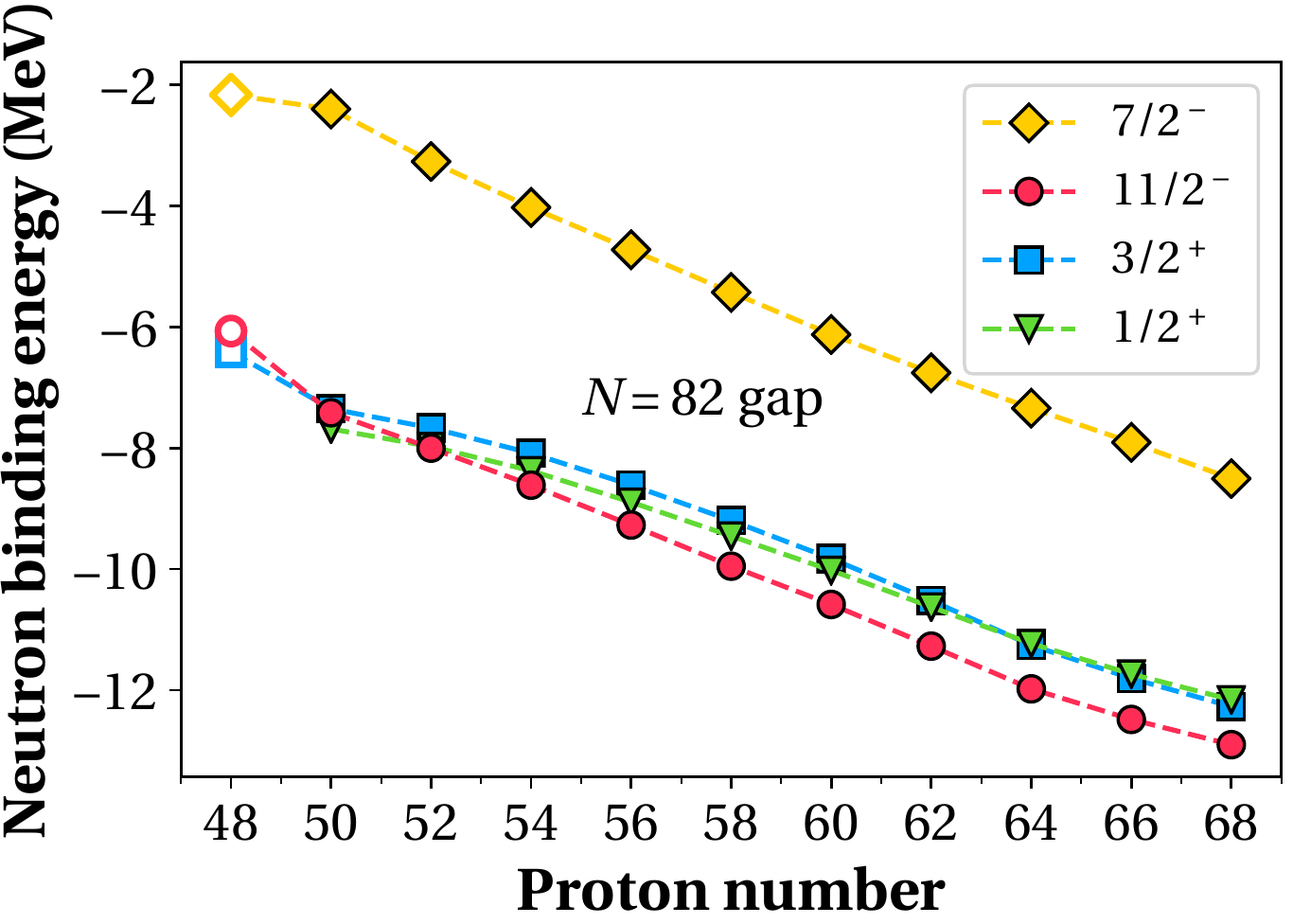}
    \caption{Neutron binding energies of the low-lying nuclear states of the $N=81$ ($J^{\pi}=1/2^+,3/2^+,11/2^-$) and $N=83$ ($J^{\pi}=7/2^-$) isotones. Experimental data are taken from \cite{NUBASE2016, ENSDF19} and this work (open symbols).}
    \label{fig:s1n-n=82}
\end{figure}

In a simple picture, the $3/2^+$ and $11/2^-$ states in $^{129}$Cd are formed by the odd neutron occupying the $d_{3/2}$ and $h_{11/2}$ orbitals, respectively, and allow probing the evolution of the two states with proton number. This is shown in Fig.~\ref{fig:s1n-n=82}, where neutron binding energies, calculated as in \cite{Sorlin2008} for the low-lying states in the even $Z$, $N=81$ and $N=83$ isotones, are plotted as a function of $Z$. For $Z=48$ they are obtained from this work. One notices the larger slope of the $11/2^-$ states, which changes more abruptly for $Z<50$, suggesting a stronger, attractive monopole proton-neutron interaction for the high-spin state.

\begin{figure}[ht!]
\centering
    \includegraphics[width=\linewidth]{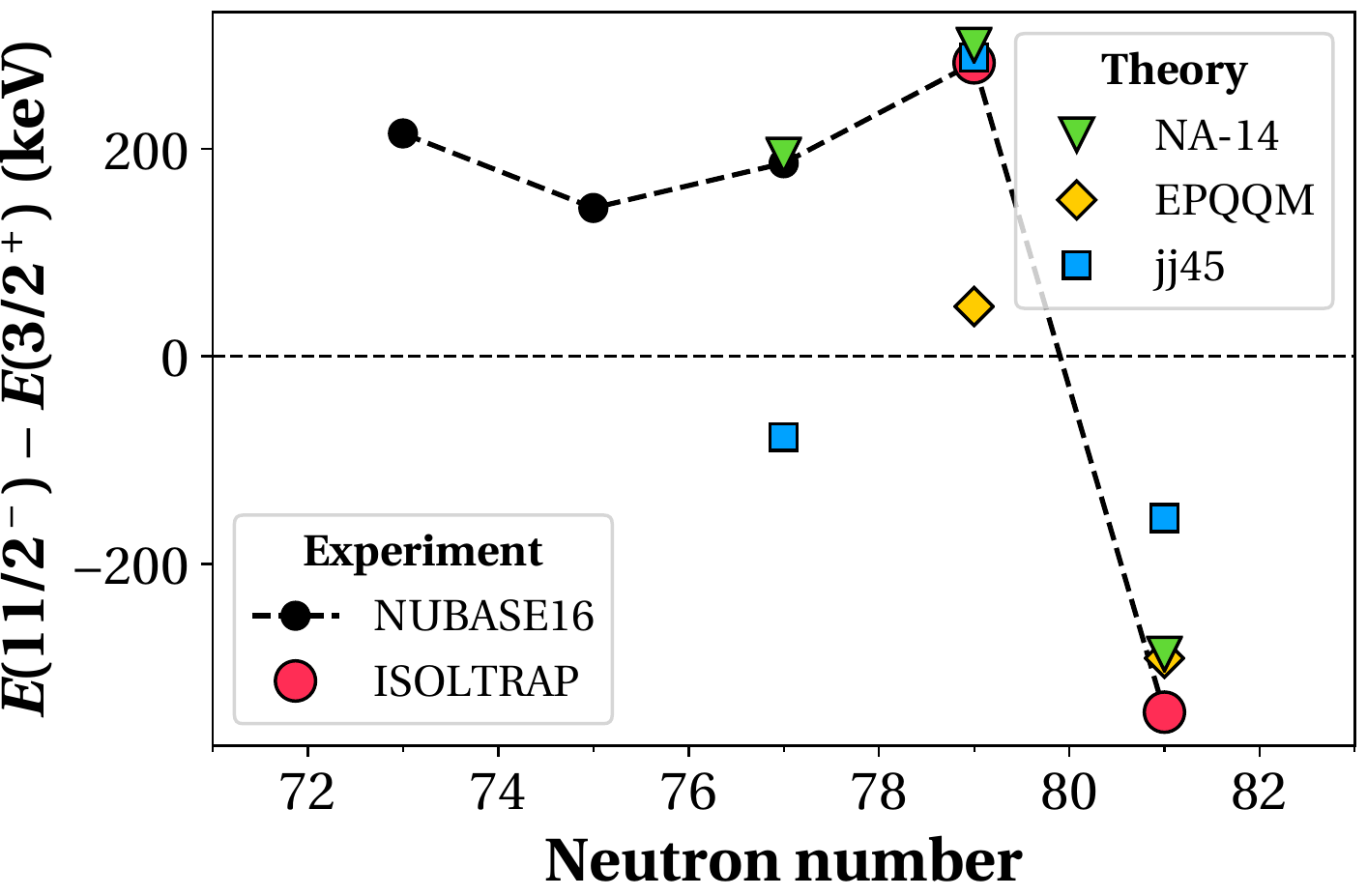}
    \caption{Energy difference between the $J=11/2^-$ and $J=3/2^+$ states in the odd cadmium isotopes. Experimental data  from \cite{NUBASE2016} and this work are compared to theoretical calculations (EPQQM  \cite{Wang2017}, NA-14 \cite{Lorenz2019,Lorenz2019b}, jj45pn \cite{Jensen95} using NUSHELLX \cite{Brown2014}).}
    \label{fig:odd-theo-vs-exp}
\end{figure}

Figure~\ref{fig:odd-theo-vs-exp} shows the difference in energy between the $3/2^+$ and $11/2^-$ states for the odd cadmium isotopes. Shell-model calculations assuming a closed $^{132}$Sn (jj45pn \cite{Jensen95,Brown2014} and NA-14 \cite{ Taprogge2014,Taprogge2015,Lorenz2019,Lorenz2019b}) or allowing cross-shell excitations (EPQQM \cite{Wang2017}) predict the $11/2^-$ state to become the ground state in $^{129}$Cd. For EPQQM, obtaining the correct prediction required enhancing the monopole interaction between the $\pi g_{9/2}$ and $\nu h_{11/2}$ orbits \cite{Wang2013}.

\begin{figure}
\begin{center}
\includegraphics[width=\columnwidth]{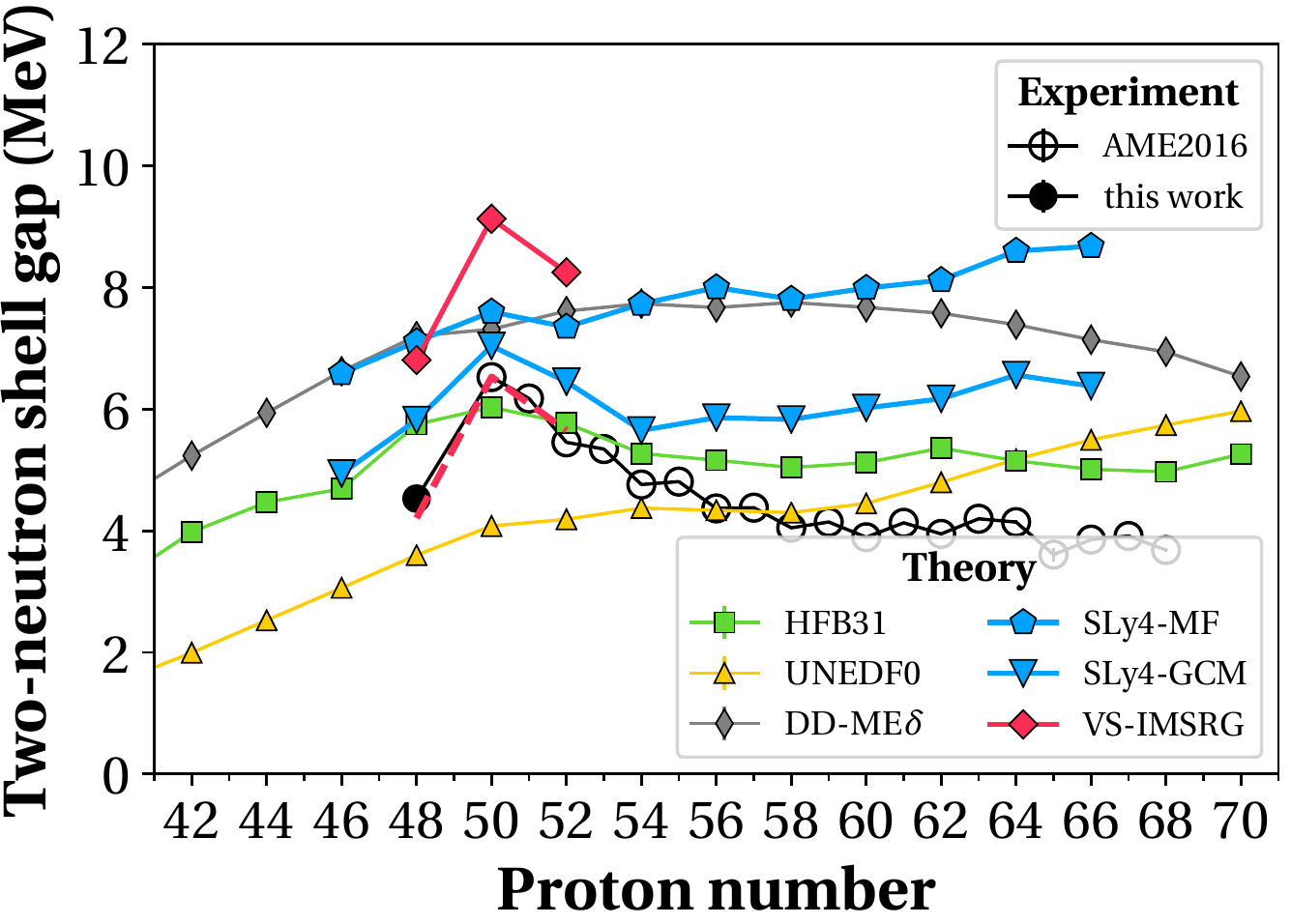}
\caption{Experimental two-neutron shell gap of the $N = 82$ isotones from the AME2016 \cite{AME2016} and this work, compared to predictions of different calculations (for details, see text). The dashed line corresponds to the VS-IMSRG results shifted to match the $Z=50$  value.}
\label{d2n}
\end{center}
\end{figure}

The mass of $^{132}$Cd allows addressing a broader range of models via the $N = 82$ two-neutron shell gap $\Delta_{2n}(Z,N) = S_{2n}(Z,N)-S_{2n}(Z,N+2)$ (where $S_{2n}$ is the two-neutron separation energy), a quantity involving only even nuclei and
the first such value below the doubly magic $^{132}$Sn. This gap is shown as a function of $Z$ in Fig.~\ref{d2n}, with the new data (full circle) revealing a peak at the proton magic number $Z=50$. This phenomenon called ``mutually enhanced magicity'' \cite{Schmidt1980,Zeldes1983} is known from other doubly-magic nuclei and was explained by a BMF calculation using the SLy4 Skyrme interaction, within a symmetry-restored generator coordinate method (GCM) \cite{Bender06, Bender08}. In this work, we show that this enhancement manifests also for $^{132}$Sn. The BMF calculations were extended to $Z = 46$ and describe the peak at $Z=50$. 
By contrast, results obtained with SLy4 just at the mean-field level (SLy4-MF) fail to reproduce the peak. It is by BMF correlations that the $N=80,84$ isotones gain binding with respect to $N = 82$, lowering the empirical shell gap, while for $Z=50$ the closed proton shell maintains the high gap value. The same failure to produce the peak in more basic mean-field calculations is also found when using other interactions. Figure~\ref{d2n} illustrates this for the nonrelativistic HFB31 \cite{Goriely2016} and UNEDF0 \cite{Kortelainen2010} Skyrme interactions and the relativistic DD-ME$\delta$ \cite{Afanasjev2016}. Calculations with HFB31 include a collective-energy correction for BMF effects, which slightly enhances $\Delta_{2n}$ around $Z=50$. While the peak is qualitatively described by BMF correlations, the size of the drop of $\Delta_{2n}$ below $Z < 50$ is not reproduced by any of these calculations.

We also present VS-IMSRG calculations of ground- and two-neutron separation energies of cadmium, tin, and tellurium isotopes across the $N=82$ shell gap. For details on the VS-IMSRG decoupling to derive the valence-space Hamiltonian, we refer to Refs. ~\cite{Tsuk12SM,Bogn14SM,Morr15Magnus,Stro16TNO,Stro17ENO,Stro2019ARNPS}. 
When this \textit{ab initio} valence-space Hamiltonian is diagonalized (here with the shell-model code ANTOINE \cite{Caurier05}) some subset of eigenvalues of the full Hamiltonian should be reproduced when no IMSRG approximations are made. In this work we use the IMSRG(2) approximation, where all induced operators are truncated at the two-body level, typically giving binding energies closer than 1\%  to full-space \textit{ab initio} results~\cite{Stro17ENO}. We begin from the 1.8/2.0(EM) chiral interaction of Refs.~\cite{Hebe11fits,Simo16unc}, used successfully throughout the medium- to heavy-mass region~\cite{Simo17SatFinNuc,Morris2018,Holt19drip}.
For heavier systems, achieving convergence with respect to the $E_{\mathrm{3max}}$ cut on $3N$ matrix elements is however a key limitation. The resulting  $\Delta_{2n}$ values are presented in Fig.~\ref{d2n}. The calculations overestimate data by almost 3~MeV, but are not fully converged with respect to the 3N matrix elements included, here up to $E_{\mathrm{3max}}=18$ excitations in a harmonic oscillator basis. In contrast, the relative trend of $\Delta_{2n}$, which is safely converged up to $\sim50$~keV, is well described.
This is illustrated by the dashed lines in Fig.~\ref{d2n}, which show the IMSRG results translated to match the $\Delta_{2n}$ value at $Z=50$.



In summary, we have measured the masses of neutron-rich cadmium isotopes and isomers across the $N = 82$ shell closure. The PI-ICR technique allowed establishing the inversion of the $11/2^-$ and $3/2^+$ states in $^{129}$Cd, showing that the $h_{11/2}$ neutron orbital is key for the evolution of the $N = 82$ shell gap towards $Z=40$. The trend of the $N=82$ shell gap was determined below $Z=50$ with the mass of $^{132}$Cd, showing a large drop, which confirms the mutually enhanced magicity of $^{132}$Sn. A BMF model reproduces the effect, but underestimates its size, whereas the VS-IMSRG approach shows an offset to experiment, but describes it qualitatively. 

\begin{acknowledgments}
V.M. and J.K. contributed equally to this work. We thank D.T. Yordanov for the helpful communication regarding the spin assignment in $^{129}$Cd and R. Stroberg for fruitful discussions on the VS-IMSRG framework. We thank the ISOLDE technical group and the ISOLDE Collaboration for their support and the excellent quality of the neutron-rich beams. We acknowledge support by the Max-Planck Society, the German Federal Ministry of Education and Research (BMBF, contracts 05P12HGCI1, 05P12HGFNE, 05P15ODCIA, 05P15HGCIA, 05P18HGCIA and 05P18RDFN1), the European Union 7th framework through ENSAR2 (contract no. 262010), the French IN2P3 and FWO Vlaanderen (Belgium). J. Karthein and A. Welker acknowledge support by a Wolfgang Gentner Ph.D. Scholarship of the BMBF (05E15CHA). W. J. Huang acknowledges the support by the China Scholarship Council (No. 201404910496).
J. Men\'endez acknowledges support from the JSPS KAKENHI Grant No.18K03639, MEXT as Priority issue on post-K computer (Elucidation of the fundamental laws and evolution of the universe), JICFuS, the CNS-RIKEN joint project for large-scale nuclear structure calculations, and the Ramón y Cajal program RYC-2017-22781 of the Spanish Ministry of Science, Innovation and Universities.
\end{acknowledgments}

%

\end{document}